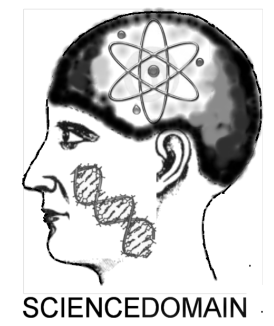

# Dynamic Model of Markets of Homogenous Non-durables

**Joachim Kaldasch[1*]**

[1]*EBC Hochschule Berlin, Alexanderplatz 1, 10178 Berlin, Germany.*

***Author's contribution***

*The sole author designed, analyzed and interpreted and prepared the manuscript.*

## ABSTRACT

A new microeconomic model is presented that aims at a description of the long-term unit sales and price evolution of homogeneous non-durable goods in polypoly markets. It merges the product lifecycle approach with the price dispersion dynamics of homogeneous goods. The model predicts a minimum critical lifetime of non-durables in order to survive. Under the condition that the supply side of the market evolves much faster than the demand side the theory suggests that unsatisfied demands are present in the first stages of the lifecycle. With the growth of production capacities these demands disappear accompanied with a logistic decrease of the mean price of the good. The model is applied to electricity as a non-durable satisfying the model condition. The presented theory allows a deeper understanding of the sales and price dynamics of non-durables.

## 1. INTRODUCTION

A microeconomic model is presented that aims at a description of the unit sales and price evolution of homogeneous non-durable goods in polypoly markets. The unit sales evolution is also called the product lifecycle of a good [1]. The idea of a product lifecycle implies that non-durables are

**Corresponding author: E-mail: joachim.kaldasch@international-business-school.de, jo@kaldasch.de;*

analogous to the life of organisms subject to characteristic stages in their sales evolution. The lifecycle of durable goods was studied intensively and a number of models turned out to be suitable to predict the evolution of durable markets [2,3]. In similarity to durables the life cycle of non-durables can be characterized by four stages, the introduction, growth, maturity and decline phase [4]. While in the introduction period a non-durable enters the market, the market penetration is related to the growth phase. This process saturates in the maturity phase. In the decline phase a current version of the non-durable is replaced by an innovation. The product lifecycle concept is applied both, as a forecasting tool and a guideline for a corporate marketing strategy [4,5]. In difference to durables the product lifecycle of non-durables is essentially determined by the evolution of the number of adopters (market diffusion) and the repurchase demand per adopter (repurchase rate). While the price is a key factor in classic microeconomic models [6,7], including the price to the lifecycle dynamics turned out to be difficult. The main approach is to treat the price as a perturbation variable in the diffusion process [8].

The presented model applies the lifecycle concept for the sales evolution of a non-durable and takes advantage from a previously published work on the price dispersion of homogeneous goods [9,10]. This work suggests that the price dispersion of a homogenous good can be approximated for a short time period by a Laplace distribution, while the mean price is governed by a Walrus equation. That means, the mean price of a good increases if there is an excess growth of demanded (required) units and decreases for the case of an excess growth of supplied (available) units [11]. Based on these ideas a dynamic approach to the product lifecycle of non-durables is established here, where the mean price is a direct consequence of the supply and demand dynamics of the market. The model predicts that the growth of the production capacities in time increases the number of available units and leads as a consequence to a decrease of the mean price. But not all demanded units can meet supplied units in the initial stages of the product lifecycle. Therefore a minimum critical lifetime of a non-durable is required in order to survive the introduction phase. Analytic relations are derived describing the evolution of the unit sales and mean price which can be compared with empirical data.

The comparison is performed here with electricity as a homogeneous good. In difference to other non-durable commodities electricity can hardly be stored. Therefore, the electricity market is based on day-ahead auctions while market clearing of the total amount of demanded and supplied units defines the spot price [12-14]. A dynamic interpretation of the long term evolution of an electricity market can be given by the application of the presented model.

The paper starts with the presentation of the dynamics of a non-durable market. Both, the unit sales and mean price evolution are established for the case that the number of supplied units growths much faster than the number of demanded units. After comparison of the model with empirical results of an electricity market the general evolution of a non-durable market is discussed followed by some conclusions.

## 2. THE MODEL

### 2.1 The Market Dynamics of a Homogeneous Good

The dynamic microeconomic model presented here is established for the case of homogeneous non-durable goods in polypoly markets. The model is based on two major presumptions:

i.) While potential buyers generate demanded (desired) units of a non-durable, suppliers provide supplied (available) product units. The purchase process of a good is treated as the meeting of demanded and supplied product units. Defining the total number of demanded units at time step *t* by $\tilde{x}(t)$ and the total number of supplied units by $\tilde{z}(t)$ purchase events must disappear if one of these numbers vanishes. Hence, the total unit sales $\tilde{y}(t)$ can be written up to the first order as a product of both variables:

$$\tilde{y}(t) \cong \eta \tilde{z}(t) \tilde{x}(t) \qquad (1)$$

with the unknown rate $\eta$. This rate characterizes the mean frequency by which the meeting of demanded and supplied product units generates successful purchase events. The parameter $\eta$ is termed meeting rate. Since $\tilde{x}(t), \tilde{z}(t), \tilde{y}(t) \geq 0$, we demand that also $\eta \geq 0$.

ii.) Both $\tilde{x}(t)$ and $\tilde{z}(t)$ are governed by conservation relations. They have the form[1]:

$$\frac{d\tilde{x}(t)}{dt} = \tilde{d}(t) - \tilde{y}(t) \tag{2}$$

and

$$\frac{d\tilde{z}(t)}{dt} = \tilde{s}(t) - \tilde{y}(t) \tag{3}$$

Eq.(2) suggests that the total number of demanded units increases with the total demand rate $\tilde{d}(t)$ which characterizes the generation rate of demanded units. It decreases by the purchase of product units with the total unit sales rate $\tilde{y}(t)$. The relation Eq. (3) states that the total number of supplied units increases with the supply of units described by the total supply rate $\tilde{s}(t)$ and decreases with the total unit sales rate $\tilde{y}(t)$. Note that these numbers can be obtained from a sum over the total number of suppliers *N(t)*:

$$\tilde{z}(t) = \sum_{k=1}^{N(t)} z_k(t); \quad \tilde{s}(t) = \sum_{k=1}^{N(t)} s_k(t); \quad \tilde{y}(t) = \sum_{k=1}^{N(t)} y_k(t) \tag{4}$$

In order to specify the time evolution of demanded and supplied product units the following processes are taken into account.

I) The evolution of the total supply rate $\tilde{s}(t)$ is governed by the growth of the production capacities of suppliers. We want to characterize this growth process with variable *γ(t)* defined by the relation between supply flow (output) and unit sales:

$$\gamma(t) = \frac{\tilde{s}(t)}{\tilde{y}(t)} - 1 \tag{5}$$

The growth of the number of supplied units is essentially determined by the mean magnitude of the variable *γ(t)*. It expresses implicitly the amount of investments in new production capacities.[2]

II) A key feature of non-durables (e.g. food) is that they have a finite mean lifetime *τ*. It is the maximum time product units of the good can be offered by the suppliers. Not purchased units disappear after this duration. Taking this effect into account Eq.(3) can be rewritten with Eq.(5) as:

$$\frac{d\tilde{z}(t)}{dt} = \gamma(t)\tilde{y}(t) - \frac{\tilde{z}(t)}{\tau} \tag{6}$$

where the first term characterizes the output evolution by the suppliers and the last term the disappearance of the current number of units with mean lifetime *τ*.

III) Also taken into account is that the number of demanded product units $\tilde{x}(t)$ may decrease not merely by purchase events, but also by a time dependent change of the demand of potential buyers. We assume that demanded units of a non-durable do not exist forever after they are generated, but have also a finite mean lifetime *Θ*.[3] Interpreting $\tilde{d}(t)$ as an effective total demand rate this effect can be included by writing:

$$\tilde{d}(t) = \tilde{d}_0(t) - \frac{\tilde{x}(t)}{\Theta} \tag{7}$$

The first term $\tilde{d}_0(t)$ expresses the general evolution of the demand rate. The second term describes the decrease of the number of demanded units $\tilde{x}(t)$ with the rate 1/Θ. For later use we introduce the maximum number of demanded units generated by the demand rate $\tilde{d}_0(t)$ by:

$$\tilde{x}_0(t) = \Theta\tilde{d}_0(t) \tag{8}$$

---

[1] *In order to establish a continuous model the integer variables in these relations are scaled by a large constant number such that they can be treated as small real numbers. We demand that this scaling leads to* $\tilde{x}(t)$, $\tilde{z}(t) \leq 1$.

[2] *This variable is also called reproduction parameter, since it characterizes the growth process of the output in the reproduction process.*

[3] *In other words, potential consumers generate a demand. But if this demand is not satisfied within an average time period Θ it disappears of its own volition. This effect can be expected in particular for non-durables.*

It characterizes the stationary number of demanded units if no purchase events occur.[4]

IV) The evolution of the demand rate $\tilde{d}_0(t)$ determines the product lifecycle of a good. The lifecycle is governed by first- and repurchase of the non-durable. First purchase is related to the spreading of the good into the market called diffusion. The diffusion process is usually described by the market penetration *n(t)* defined by:

$$n(t) = \frac{N_A(t)}{M_A} \tag{9}$$

where $N_A(t)$ is the cumulative number of adopters and $M_A$ the market potential, i.e. the number of all possible adopters of a good[5]. The evolution of the number of adopters can be written as a conservation relation of the form:

$$\frac{dn(t)}{dt} = \varphi(t)n'(t) - \theta(t)n(t) \tag{10}$$

The first term indicates the generation of new adopters. It is proportional to the generation rate of adopters *φ(t)* and the number of potential adopter *n'(t)* not yet adopted the good. This number is determined by the difference between the maximum number of adopters $n_{max}$ and the actual number *n(t)*:

$$n'(t) = n_{\max} - n(t) \tag{11}$$

The second term in Eq.(10) indicates the decline of *n(t)* with a decline rate *θ(t)*. In the introduction, growth and maturity phase of the lifecycle the decline rate is $\theta(t) \approx 0$. In the decline phase we demand that $\varphi(t)=0$, $\theta(t) \neq 0$.

Expanding the generation rate *φ(t)* as a function of the number of adopters we obtain up to the first order:

$$\varphi(t) \cong \mathrm{A} + \mathrm{B}n(t) + ... \tag{12}$$

with constant coefficients $A,B>0$. Inserting this relation into Eq.(10) yields for $\theta(t)=0$ a standard approach to describe the diffusion processes of goods known as the Bass model [15]. It has the form:

$$\frac{dn(t)}{dt} = \mathrm{A}(n_{\max} - n(t)) + \mathrm{B}n(t)(n_{\max} - n(t)) \tag{13}$$

The first term is interpreted as spontaneous purchase by potential adopters, where *A* is the so-called innovation rate. The second term is due to social learning, where *n(t)* increases with an imitation rate *B*. Formally Eq.(13) can be used to distinguish between the characteristic stages of the lifecycle by the dominant process generating adopters. In the introduction phase the first term dominates over the second. That means in this phase spontaneous purchase governs the diffusion process. Note that the introduction phase is often dominated by a single supplier (monopoly). It is therefore not in the focus of this model. When the second term in Eq.(13) dominates over the first the good is in the growth phase. In this period the main adopter generation process is social learning. In the maturity phase the number of adopters approaches its maximum number $n(t) \approx n_{max}$.

The market penetration suggested by the Bass diffusion model has the form:

$$n(t) = \frac{1 - e^{-(A+B)t}}{1 + \frac{B}{A} e^{-(A+B)t}} n_{\max} \tag{14}$$

While first purchase (compared to repurchase) plays a crucial role for durable goods, for non-durables the impact of first purchase events on the unit sales can be neglected. Repurchase events can be regarded to be proportional to the current number of adopters *n(t)*. The total repurchase sales of a non-durable can therefore be modelled as the product of *n(t)* and a time dependent repurchase rate *ξ(t)* characterizing the average number of repurchased units per unit time and adopter. The total unit sales of the non-durable can therefore be described by:

$$\tilde{y}(t) = \xi(t)n(t) \tag{15}$$

The repurchase rate *ξ(t)* is written here as the sum of a time independent constant and a time dependent contribution:

---

[4] *The variable $\tilde{x}_0(t)$ can be obtained from the stationary solution of Eq.(2) using Eq.(7) and setting $\tilde{y}(t)=0$.*

[5] *The evolution of $M_A$ is neglected here.*

$$\xi(t) = \xi_0 + \xi_r(t) \tag{16}$$

The constant $\xi_0$ indicates a minimum average number of units per purchase event. The time dependent repurchase rate $\xi_r(t)$ takes the growth of the repurchase rate with the evolution of the considered economy into account. In order to keep the model simple we assume that the repurchase rate $\xi_r(t)$ cannot grow up to infinity and demand that $\xi_r(t)$ approaches a maximum magnitude $b_\xi$ after sufficient time. Such a constraint growth can be described by a logistic differential equation of the form:

$$\frac{d\xi_r(t)}{dt} = a_\xi \xi_r(t)\left(1 - \frac{\xi_r(t)}{b_\xi}\right) \tag{17}$$

where $a_\xi$ and $b_\xi$ are free parameters. The repurchase rate $\xi_r(t)$ becomes:

$$\xi_r(t) = \frac{b_\xi}{1 + C_\xi e^{-a_\xi t}} \tag{18}$$

with the unknown constant $C_\xi$.

## 2.2 The Sales Evolution of a Homogeneous Good

The total numbers of demand units $\tilde{x}(t)$ and supplied units $\tilde{z}(t)$ fluctuate considerable in time. If time-dependent variations of these variables are of the same order the sales evolution can hardly be predicted. In order to keep the model simple the theory is confined here to markets dominated by the supply side dynamics in a considered time interval $\Delta t$. Thus we confine the model to free markets where the number of supplied units evolves much faster than the number of demanded units:

$$\frac{d\tilde{x}(t)}{dt} << \frac{d\tilde{z}(t)}{dt} \tag{19}$$

If $\tilde{x}(t)$ evolves sufficiently slowly in comparison with $\tilde{z}(t)$ we can approximately write:[6]

$$d\tilde{x}(t)/dt \cong 0 \tag{20}$$

[6] *It is the so-called adiabatic approximation.*

and directly obtain from Eq.(2):

$$\tilde{y}(t) \cong \tilde{d}_0(t) - \frac{1}{\Theta}\tilde{x}(t) \tag{21}$$

where we used Eq.(7). Hence, the purchase rate is equal to the generation rate of demanded units $\tilde{d}_0(t)$ diminished by the rate $\tilde{x}(t)/\Theta$. Applying Eq.(1) and Eq.(8) we further get for the total number of demanded units:

$$\tilde{x}(t) = \frac{\tilde{x}_0(t)}{1 + \Theta\eta\tilde{z}(t)} \cong \tilde{x}_0(t) - \Theta\eta\tilde{x}_0(t)\tilde{z}(t) \tag{22}$$

where we used in the approximation that $\tilde{z}(t) \leq 1$. Applying this result in Eq.(1), Eq.(6) turns into:

$$\frac{d\tilde{z}(t)}{dt} = \alpha\tilde{z}(t) - \Theta\eta\alpha'\tilde{z}(t)^2 \tag{23}$$

with the time averaged parameters:

$$\alpha' = \frac{1}{\Delta t}\int_t^{t+\Delta t} \eta\gamma(t')\tilde{x}_0(t')dt'; \qquad \alpha = \alpha' - \frac{1}{\tau} \tag{24}$$

Note that for a sufficiently long lifetime $\tau$ of the good is $\alpha \approx \alpha'$. Since Eq.(23) is a logistic differential equation the evolution of the number of supplied units can be given by:

$$\tilde{z}(t) = \frac{\tilde{z}_0}{1 + C_z e^{-\alpha t}} \tag{25}$$

where $C_z$ is a constant and the maximum number of supplied units is:

$$\tilde{z}_0 = \frac{\alpha}{\alpha'\eta\Theta} \tag{26}$$

The sales evolution of a non-durable depends therefore mainly on the magnitude of $\alpha$. For $\alpha<0$, there is on time average a supply shortage of product units. In this case the number of available units disappears in time $\tilde{z}(t) \rightarrow 0$. Eq.(24) suggests that this may happen if the lifetime $\tau$ of the good is very short, since then there is not sufficient time for potential buyers to purchase available units during the time they are offered. The critical lifetime $\tau_c$ can be estimated from Eq.(24) by setting $\alpha=0$. It leads to $\tau_c \geq 1/(\eta\gamma x_0)$. Thus, a non-durable good must have a high mean reproduction parameter $\gamma$ (high output compared to unit sales), a high meeting rate $\eta$ (high availability) and a sufficient number of potential consumers $x_0$ (high demand) in order to survive the introduction phase of the product lifecycle.

We want to confine the discussion here to $\alpha>0$. In this case the supply side extends the production capacities and the number of available units increases in time until $\tilde{z}(t)=\tilde{z}_0$. The total unit sales Eq.(1) evolve with Eq.(22) and Eq.(25) as:

$$\tilde{y}(t)=\frac{\alpha}{\alpha'}\frac{\tilde{d}_0(t)}{1+C_z e^{-\alpha t}} \cong \frac{\tilde{d}_0(t)}{1+C_z e^{-\alpha t}} \quad (27)$$

This relation suggests that at introduction of the good $(t=0)$ the unit sales are smaller than the generation rate of demanded units. That means that there are unsatisfied demands in the initial stages of the lifecycle. The increase of the production capacities for $\alpha>0$, however, decreases this amount. As a consequence the total unit sales $\tilde{y}(t)$ increase in time until $\tilde{z}(t)=\tilde{z}_0$ where the sales evolution is completely determined by the demand rate $\tilde{d}_0(t)$. Inserting Eq.(26) in Eq.(22) we can conclude that in this state the number of demanded units is proportional to the maximum number of demanded units $\tilde{x}(t)=\tilde{x}_0(t)(1+\alpha/\alpha') \approx \tilde{x}_0(t)/2$. And since $\tilde{z}_0$ is a constant in this saturated state, we can further determine from Eq.(6) and the condition $d\tilde{z}_0/dt=0$ that $\tilde{x}_0(t)=(\eta\tau\gamma(t))^{-1}$.

The model suggests therefore that the demand for a good is much higher in the initial stages of the lifecycle than expressed by the realized unit sales $\tilde{y}(t)$. The demand rate $\tilde{d}_0(t)$ is, however, not directly available. The only empirically observable variable is $\tilde{y}(t)$. But as will become clear in the next chapter, the unsatisfied demands have an impact on the price of a good.

In particular the evolution of the number of available units $\tilde{z}(t)$ plays a crucial role in the determination of the mean price of a non-durable.

### 2.3 The Price Evolution of a Homogeneous Good

We want to study the price evolution of a homogeneous non-durable in a polypoly market as a result of the sales dynamics established in the previous chapter. For this purpose presumption i) is generalized. We assume that the number of purchase events in a given price interval $p$ and $p+dp$ must disappear if either the number of demanded units at this price $x(t,p)$ or the number of supplied units $z(t,p)$ vanishes. Hence, the unit sales at a given price $y(t,p)$ can be approximated in similarity to Eq.(1) as:

$$y(t,p) \cong \eta z(t,p)x(t,p) \quad (28)$$

where the meeting rate $\eta$ is treated as price independent. The price dispersion is determined by the relative abundance of purchase events in a price interval given by:

$$P(t,p)=\frac{1}{\tilde{y}(t)}y(t,p) \quad (29)$$

The mean price is defined by:

$$\mu(t)=\int_0^\infty P(t,p)p\,dp \quad (30)$$

As shown in a previous work and known from empirical investigations the price dispersion of homogeneous goods can be approximately described for short time horizons by a Laplace distribution [10]:

$$P(p) \cong \frac{1}{2\sigma}e^{-\frac{|p-\mu|}{\sigma}} \quad (31)$$

where the standard deviation $\sigma$ is a function of the mean price $\sigma=\sqrt{2}(\mu-\mu_m)$. The minimum price $\mu_m \geq 0$ indicates a price below which the production of the good is not profitable.[7] Further

[7] *The minimum mean price $\mu_m$ is considered to be the lowest price that can be offered with the applied production*

shown is that the mean price is governed by a Walrus equation relation of the form [10]:

$$\frac{1}{\mu(t)-\mu_m}\frac{d\mu}{dt} = H(t)\left(\frac{d\tilde{x}}{dt}-\frac{d\tilde{z}}{dt}\right) \quad (32)$$

where:

$$H(t) \cong \frac{1}{\sqrt{2\tilde{x}(t)}} \quad (33)$$

Applying Eq.(2) and Eq.(3) this relation states that an excess total demand rate $\tilde{d}(t)$ increases and an excess total supply rate $\tilde{s}(t)$ decreases the mean price in time.

Eq.(32) can be used to determine the evolution of the mean price of a non-durable homogeneous good. Taking advantage from Eq.(20) we obtain:

$$\frac{1}{\mu(t)-\mu_m}\frac{d\mu(t)}{dt} \cong -H\frac{d\tilde{z}(t)}{dt} \quad (34)$$

where $\tilde{x}(t)$ is approximated by the constant $\tilde{x}_0$. Applying Eq.(23) we further get:

$$\frac{d\mu(t)}{dt} \cong -H\alpha(t)\tilde{z}(t)\big(\mu(t)-\mu_m\big) \quad (35)$$

while higher order terms in $\tilde{z}(t)$ are neglected.

The evolution of the mean price depends on the sign of *α(t).* For *α(t)<0*, the supply shortage generates an exponential increase of the price until $\tilde{z}(t)=0$ (inflation). For *α(t)>0*, however, the mean price declines as a result of the excess supply (deflation). It approaches a stationary state given by either $\mu=\mu_m$ or $\tilde{z}=\tilde{z}_0$. In the first case Eq.(35) reduces for $\tilde{z}(t)<\tilde{z}_0$ to:

$$\frac{d\mu(t)}{dt} = -a(t)\big(\mu(t)-\mu_m\big) \quad (36)$$

with a mean price decline rate $a(t) \cong H\tilde{z}(t)\alpha(t)$. The mean price evolution has then the form:

$$\mu(t) \cong \mu_0 e^{-\int a(t)dt} + \mu_m \quad (37)$$

where $\mu(0)=\mu_0+\mu_m$. The main trend of the mean price is in this case an exponential decline towards the stationary minimum price $\mu_m$. However, because the standard deviation of the price dispersion disappears at $\mu_m$ this case ends up with a monopoly market. Since we focus on polypoly markets here, this case is not further considered.

For $\mu(t)>\mu_m$ Eq.(34) can be rewritten as:

$$\int\frac{d\mu(t)}{\mu(t)-\mu_m} \cong -H\int d\tilde{z}(t) \quad (38)$$

and we obtain:

$$\mu(t) = \mu_0 e^{-H\tilde{z}(t)} + \mu_m \cong \mu_0\big(1-H\tilde{z}(t)\big)+\mu_m \quad (39)$$

while we used in the approximation that $\tilde{z}(t) \le 1$. The mean price declines with the function $\tilde{z}(t)$ (Eq.(25)) governed by a logistic law. Since $\tilde{z}(t) \to \tilde{z}_0$ the mean price approaches a floor price $\mu_f>\mu_m$ given by:

$$\mu_f = \mu_0\exp(-H\tilde{z}_0)+\mu_m \quad (40)$$

The increase of the total output is therefore directly related to the evolution of *μ(t)*. When $\tilde{z}(t) \to \tilde{z}_0$, the mean price decreases towards the floor price $\mu_f$ which is a stationary market without unsatisfied demands (the unit sales are equal to the generation rate of demanded units). Note that the previous work [10] suggests that the symmetry of the price dispersion implies $\tilde{z}_0 \approx \tilde{x}_0$ [9]. When the market is at $\mu_f$ the mean lifetime of demanded units can therefore approximately obtained from:

$$\Theta \cong \sqrt{\frac{1}{\eta d_0}} \quad (41)$$

where we used Eq.(8) and Eq.(26).

*technology. The applied technology is assumed to evolve very slowly such that $\mu_m$ can be treated as constant.*

## 3. COMPARISON WITH THE EMPIRICAL RESULTS

The presented theory suggests the following characteristics of the evolution of a polypoly market of homogeneous non-durables when capacities growth sufficiently fast:

1. The price dispersion *P(p)* of homogenous non-durables are given for short time periods by Eq.(31) and the mean price *μ(t)* decreases over long time periods according to the logistic law Eq.(39).
2. The evolution of the total unit sales $\tilde{y}(t)$ is essentially determined by the repurchase of the non-durable Eq.(15), which is governed by the market penetration Eq.(14) and the growth of the repurchase rate Eq.(18).

For a comparison of these statements with empirical data we want to take advantage from investigations on electricity as a non-durable homogeneous good. The electricity market is a polypoly day-ahead market of power generating companies on the supply side, utilities and large industrial consumers on the demand side. [8] Details of this market are discussed for example by Geman [16] not further outlined here.

The first assertion suggests that the price dispersion of a homogenous good can be approximated by a Laplace distribution for short time periods. As an example the central part of a filtered price dispersion of the Nordpool electricity market [17] is shown in Fig. 1 and fitted with Eq.(31). The good coincidence indicates the applicability of the presented model to electricity markets as a homogeneous good.

The model further suggests that the long term evolution of the mean price for a fast growing output can be described by Eq.(39). In order to compare this statement with empirical investigations, we take advantage from historical data of the US non-industrial electricity price of the last century presented by Ayres and Warr [18]. They are displayed in Fig. 2 and taken as a measure of the mean price *μ(t)* indicated by dots in this graph. Note that in the introduction phase the electricity market was separated into different grids of single suppliers (monopolies). The electricity act of 1926 led to the setting up of the national electricity grid. In order to satisfy the model conditions the fit of Eq.(39) with empirical price data is therefore confined here to the time period after 1926.

The theory suggests that if suppliers expand their capacities faster than the rise of demanded units the freely available amount of units increases with time. This expansion process causes a decrease of the mean price as can be found in the empirical price evolution. The fit of the logistic mean price decline with Eq.(39) suggests a time evolution of $\tilde{z}(t)$ (Eq.(25)) as displayed in the insert of Fig. 2 while $\tilde{z}_0 = 1$. The mean price converges to a constant floor price $\mu_f \approx 6$ *Cent/kWH*. Unfortunately the available data do not allow the specification of the lifetime *Θ* by the application of Eq.(41).

Also displayed in Fig. 2 is the evolution of the total unit sales $\tilde{y}(t)$. The empirical data of the electricity lifecycle are given in terms of an index which characterizes the electricity unit sales in relation to the *1902* magnitude [18]. The second assertion suggests that the unit sales are determined by the repurchase of the non-durable. Repurchase is in this model proportional to the market penetration *n(t)* shown in Fig. 3. The empirical diffusion process of electricity follows the expected S-curve (solid line) [19]. The fit of the empirical data with the Bass model Eq.(14) (dashed line) ignores perturbations of the diffusion process from the expected mean growth in the roaring twenties and the following depression (since $\theta=0$ in this period). A fit of the logistic growth of the repurchase rate *ξ(t)* applying the market penetration in Fig. 3 leads to the total unit sales $\tilde{y}(t)$ (Eq.(15)) indicated by the dashed line in Fig. 1. The data suggest that the growth phase of the life cycle ended around 1970, since the market penetration was completed there. In the maturity phase the main growth of electricity demand is due to the increase of the repurchase rate *ξ(t)*.

From the application of the model to the electricity market we can conclude that the highest growth of capacities took place after World War II around 1950, as can be seen by the rapid increase of available units $\tilde{z}(t)$ in the insert of Fig. 2. It generates not only a considerable decrease of the mean price for electricity, but leads also to the convergence of the market penetration *n(t)* to its maximum magnitude.

[8] *The lifetime τ is therefore 1 day for this good.*

## 4. DISCUSSION

For the case that a market is dominated by a fast growing supply side the presented model predicts the unit sales and mean price evolution of a non-durable. It suggests that the unit sales are governed on the one hand by the spreading (diffusion) of the good into the market, described here by the Bass model. On the other hand they are determined by the amount of repurchased units caused by the growth of the considered economy. For electricity as a non-durable the unit sales and mean price evolution is schematically illustrated in Fig. 4. Shown is the price dispersion *P(p)* at a time step in the growth phase of the lifecycle indicated by the dotted line (right peak). Generally the price dispersion has the form of a Laplace distribution as found empirically (see Fig. 1). It is a consequence of the meeting of the cumulated number of demanded units *x(p,t)* (demand curve) with the cumulated number of supplied units *z(p,t)* (supply curve) also shown in Fig. 4. The chance that demanded and supplied units meet at a given price *p* has its maximum at mean price *μ(t)* where the functions *x(p,t)* and *z(p,t)* have maximum overlap [10]. The supply function *z(p,t)* is governed by the costs per unit for the generation of electricity, which depends on the applied production technology. The smallest prices are offered by nuclear power stations while the highest prices are generally demanded by gas power stations indicated in the figure.

The main idea to understand the price evolution of the electricity market is that there are unsatisfied demands in the growth phase of the product lifecycle. For the case *α>0*, the production capacities increase faster than the unit sales. As a result the number of available units $\tilde{z}(t)$ growth in time. This growth leads to the decline of the mean price governed by a logistic law as displayed in the insert of Fig. 4. When $\tilde{z}(t)$ approaches $\tilde{z}_0$ unsatisfied demands disappear and the mean price *μ(t)* approaches the constant floor price $\mu_f$. Since the variance of the price dispersion depends explicitly on the mean price, *P(p)* becomes a sharp distribution around $\mu_f$ in this state (left peak).

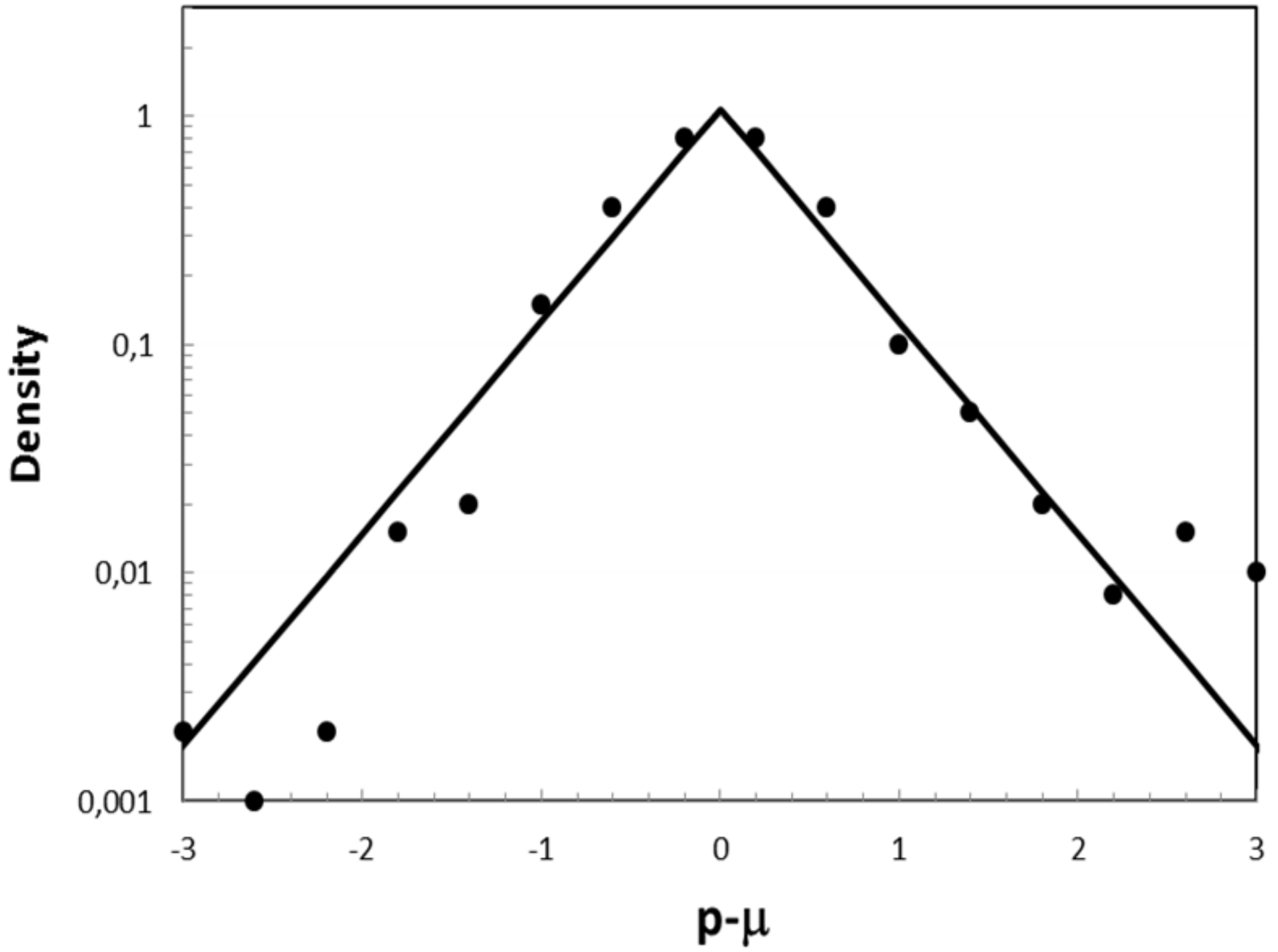

**Fig. 1. The price distribution *P(p)* of the Noordpool electricity market (at 1 am) around a scaled mean price *μ=0*. The dots indicate empirical data (Sapio 2004). The solid line is a fit of Eq.(31) with *σ=0.468***

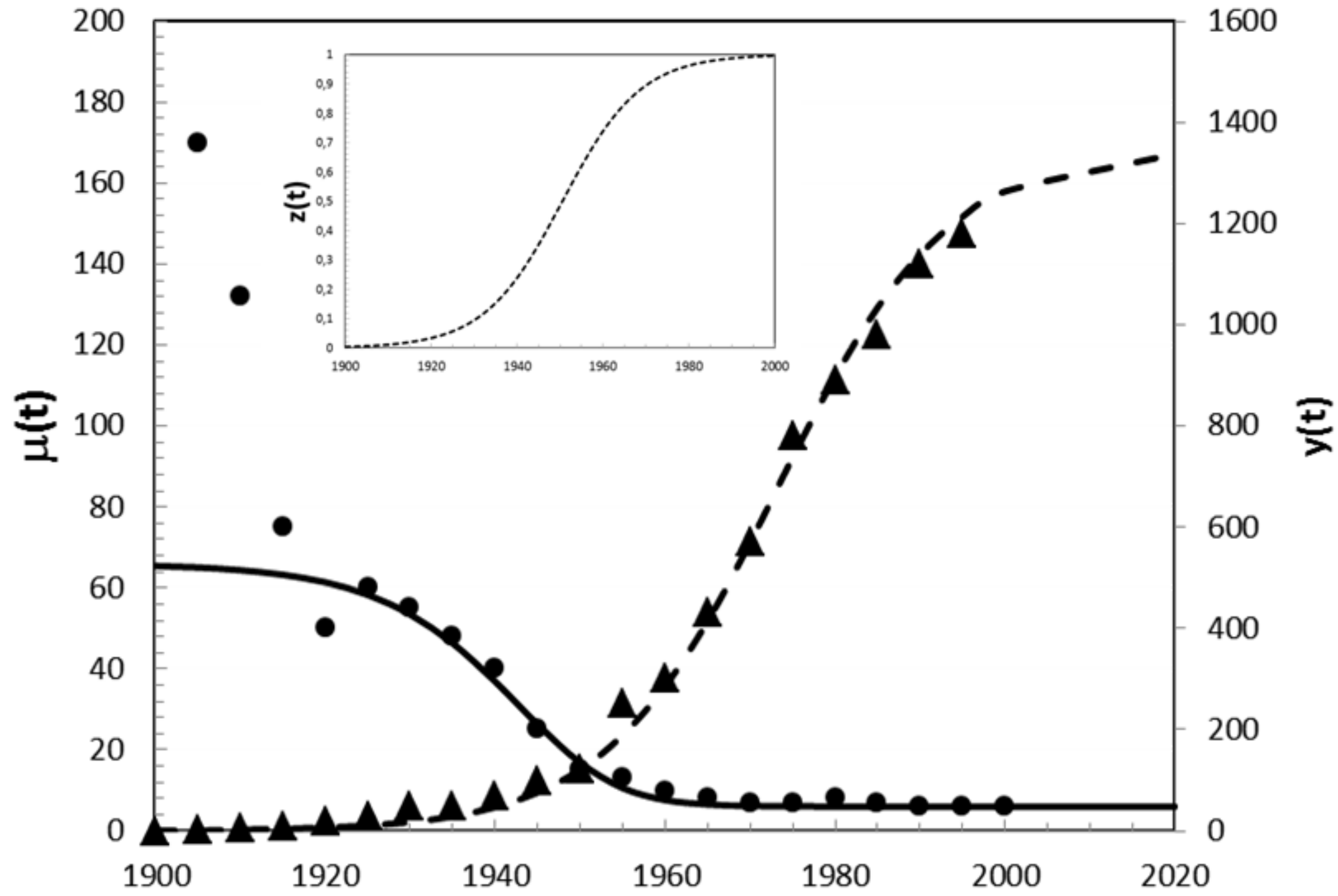

**Fig. 2. Evolution of the US non-industrial electricity price $\mu(t)$ in cent per kWh (constant 1992 US$). The dots indicate empirical data [18]. The solid line is a fit of Eq.(39) with $\mu_m$=*1 cent/kWh*, $\mu_0$= *70 cent/ kWh*, $H\tilde{z}_0$=*2.8.* Displayed in the insert is the function $\tilde{z}(t)$ given by Eq.(25) with $\tilde{z}_0 = 1$, *α=0.11 per year* and $C_z$=*260.* Also shown are the historical unit sales $\tilde{y}(t)$ (triangles) in terms of an index indicating the electricity demand in relation to the magnitude in *1902*. The dashed line is a fit of the long term unit sales given by Eq.(15), where we used $n(t)$ from Fig. 3 and $\xi_0$=*0.2* $a_\xi$ =*0.1 per year,* $b_\xi$ =*30* and $C_\xi$=*1500***

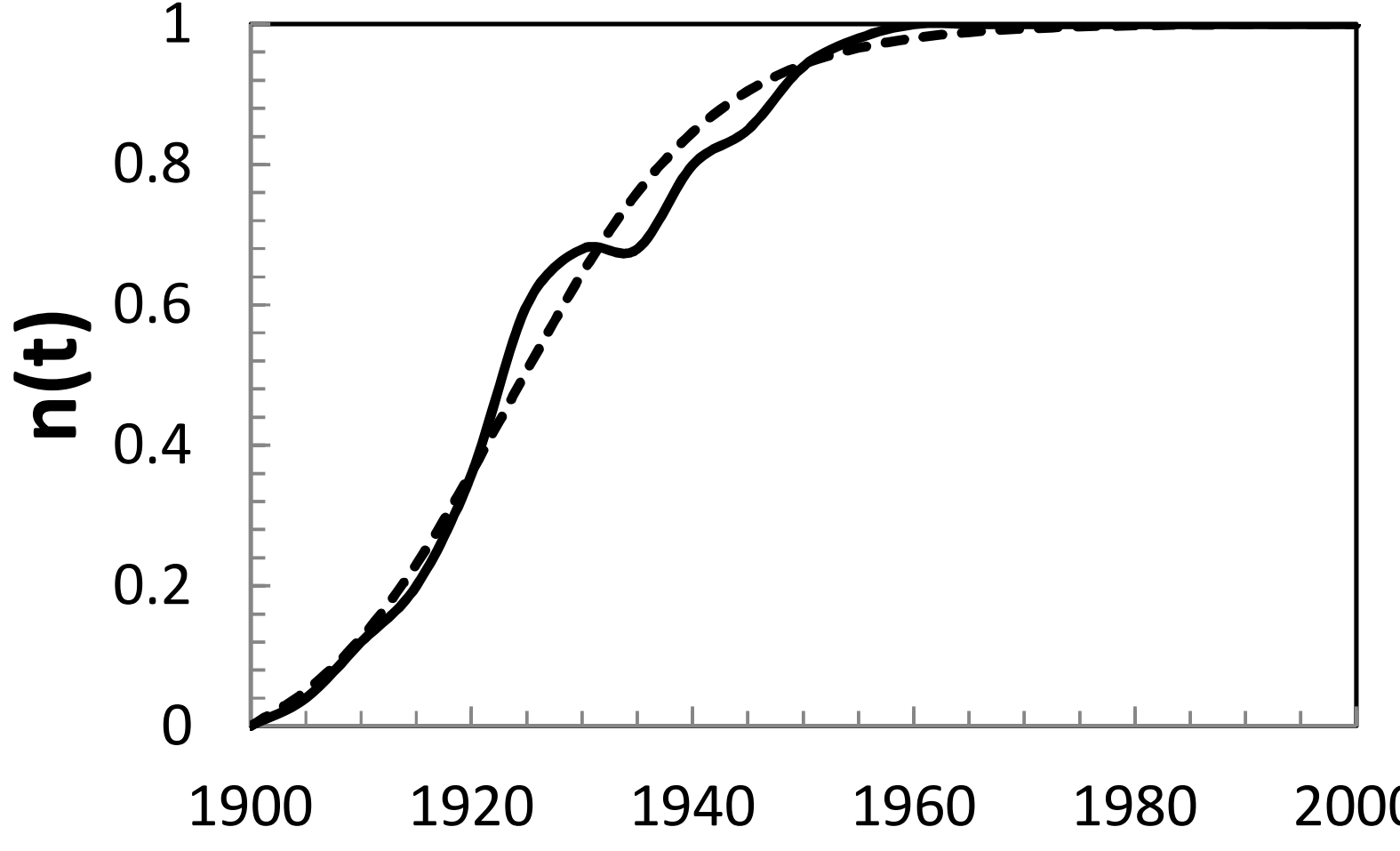

**Fig. 3. The solid line indicates empirical data of the electricity penetration $n(t)$ in the USA [19]. The dashed line is a fit with the Bass model Eq.(14) using $n_{max}$=*1, A=0.008, B=0.1***

Fig. 4. Schematically displayed are the price dependent demand and supply curves *x(p,t)*, *z(p,t)* (solid lines) and the price distributions *P(p)* (dotted lines) of the electricity market at two different time steps. The supply curve *z(p,t)* depends on the costs per unit governed by the applied generation technology. The model suggests that with the increase of the number of available units $\tilde{z}(t) \rightarrow \tilde{z}_0$ the mean price *μ(t)* declines until $\mu(t) \rightarrow \mu_f$ (see insert). As a consequence the price dispersion for $\mu(t)=\mu_f$ and $\tilde{z}(t) = \tilde{z}_0$ becomes a narrow peak

## 5. CONCLUSION

The presented microeconomic theory is aimed at a deeper understanding of the market dynamics of homogeneous non-durable goods. It is based on a combination of the product lifecycle concept with the dynamics of the price evolution of homogeneous goods. The dynamic theory suggests that non-durables must have a minimum critical lifetime in order to survive the introduction phase. Its magnitude depends essentially on the availability and demand for the good. Due to limited production capacities in the first stages of the lifecycle the presented approach predicts the existence of unsatisfied demands. With increasing output they decrease in time associated with a logistic decline of the mean price. The unit sales of non-durables, are in difference to durables [20] essentially determined by the repurchase of the good. Repurchase is modelled here as proportional to the current number of adopters and a repurchase rate governed by a logistic growth. Although the model is applied here merely to the electricity market it can be expected that it also applies to other fast growing non-durable markets, if it is not dominated by speculation.

## COMPETING INTERESTS

Author has declared that no competing interests exist.

## REFERENCES

1. Saaksvuori A, Immonen A. Product lifecycle management. Springer Verlag; 2008.
2. Mahajan V, Muller E, Wind Y. New-product diffusion models. Springer Verlag; 2000.

3. Rogers EM. Diffusion of innovations. Simon and Schuster, New York; 2003.
4. Chitale AK, Gupta R. Product policy and brand management. Prentice-Hall of India Pvt. Ltd.; 2010.
5. Hollensen S. Marketing management. Pearson Education Ltd.; 2010.
6. Varian HR. Intermediate microeconomics – A modern approach. W.W. Northern & Company Inc. New York; 2006.
7. Cantner U, Krueger JJ, Soellner R. Product quality, product price, and share dynamics in the German compact car market. Jena Economic Research Papers. 2010; 24.
8. Chandrasekaran D, Tellis GJ. A critical review of marketing research on diffusion of new products. Rev. Marketing Research. 2007;3:39-80.
9. Kaldasch J. Evolutionary model of stock markets. Physica A. 2014;415:449-462.
10. Kaldasch J. Dynamic model of the price dispersion of homogeneous goods. British Journal of Economics, Management & Trade. 2015;8(2): 120-131.
11. Zhang WB. Synergetic economics. Springer Verlag Berlin; 1991.
12. Bunn DW. Modelling prices in competitive electricity markets. Wiley Finance; 2004.
13. Sioshansi FP. Evolution of global electricity markets. Academic Press; 2013.
14. Biggar DR, Hesamzadeh MR. The economics of electricity markets. John Wiley & Sons Ltd.; 2014.
15. Bass FM. A new product growth model for consumer durables. Management Science. 1969;15:215-227.
16. Geman H. Commodities and commodity derivatives: Modelling and pricing for agriculturals, metals, and energy. John Wiley & Sons Ltd.; 2005.
17. Sapio S. An empirically based model of the supply schedule in day-ahead electricity markets. LEM Working Paper Series. 2008;13.
18. Ayres RU, Warr B. Accounting for growth: The role of physical work. Structural Change & Economic Dynamics. 2005;16: 181-209.
19. Felton N. You are what you spend. New York Times, 10. Feb. 2008. Available:http://visualeconsite.s3.amazonaws.com/wp-content/uploads/2008/02/history-of-products.gif (Accessed 8 June 2015).
20. Kaldasch J. Evolutionary model of an anonymous consumer durable market. Physica A. 2011;390:2692-2715.